\documentclass{KapProc} 
\usepackage{epsf}

\kluwerbib

 \normallatexbib 
\begin{document}
\newcommand{\hmp}{h^{-1}Mpc}      
\newcommand{\Ga}{\Gamma}     
\newcommand{\Om}{\Omega}     
\newcommand{\de}{\delta}     
\newcommand{\al}{\alpha}     
\newcommand{\si}{\sigma}     
\newcommand{\bx}{{\bf x}}     
\newcommand{\lam}{\lambda}     
\newcommand{\lan}{\langle}     
\newcommand{\ran}{\rangle}     
\newcommand{\La}{\Lambda}     
\newcommand{\bm}{\boldmath}     
\newcommand{\be}{\begin{equation}}     
\newcommand{\ee}{\end{equation}}     
\newcommand{\bea}{\begin{eqnarray}}     
\newcommand{\eea}{\end{eqnarray}}     
\newcommand{\ra}{\rightarrow}     
\newcommand{\bef}{\begin{figure}}     
\newcommand{\eef}{\end{figure}}     
\newcommand{\Mpc}{{\rm Mpc}}     
       
\newcommand{\veps}{\varepsilon}      
\def\spose#1{\hbox to 0pt{#1\hss}}      
\def\ltapprox{\mathrel{\spose{\lower 3pt\hbox{$\mathchar"218$}}      
 \raise 2.0pt\hbox{$\mathchar"13C$}}}      
\def\gtapprox{\mathrel{\spose{\lower 3pt\hbox{$\mathchar"218$}}      
 \raise 2.0pt\hbox{$\mathchar"13E$}}}      
\def\inapprox{\mathrel{\spose{\lower 3pt\hbox{$\mathchar"218$}}      
 \raise 2.0pt\hbox{$\mathchar"232$}}}      
\articletitle{Complexity in Cosmology}

\articlesubtitle{Statistical properties of galaxy large scale structures}

\author{Francesco Sylos Labini}
\affil{Dept.~de Physique Theorique, Universite de Geneve\\
24, Quai E. Ansermet, CH-1211 Geneve, Switzerland}
\email{sylos@amorgos.unige.ch}

\author{Luciano Pietronero}
\affil{INFM Sezione Roma 1 \& 
Dip. di Fisica, Universita' ``La Sapienza'', \\
P.le A. Moro 2 I-00185 Roma, Italy. }
\email{luciano@pil.phys.uniroma1.it}

\begin{abstract}
The question of the nature of galaxy clustering and 
the possible homogeneity of galaxy distribution is one of
the fundamental problem of cosmology. It is well established that
galaxy structures are characterized, up to a certain
scale, by fractal properties.
The possible crossover to homogeneity is instead still 
matter of debate. However, independently on the specific
value of the homogeneity scale, the fractal nature of galaxy clustering
requires new methods and theoretical concepts
developed in the area of statistical 
physics and complexity.  In this lecture we
discuss a  new perspective
on the problem of cosmological structures formation,
both from the experimental and the theoretical points of view. 
This new perspective leads to a very interesting and constructive 
interaction
between the fields of cosmic structures, statistical physics and 
complexity with very challenging open problems which we also discuss.
\end{abstract}


\section{Introduction}
The large amount of new data which is accumulating for galaxy distribution 
and for the cosmic microwave background radiation (CMBR) 
calls for a characterization of structures and correlations 
in terms of concepts developed in the area of statistical 
physics and complexity. The two observations, however, appear quite different.
On one hand the CMBR is extremely isotropic and the small amplitude, possible
Gaussian, approach to characterize its fluctuations 
seems rather adequate. On the other hand 
galaxies show highly structured patterns, with fractal like properties 
and for which the definition of background average density
is still a matter of debate. Up to now our activity has been focused
mostly on galaxy distribution. We have shown the importance
of fractal properties and their implications 
on the statistical methods and on the theoretical framework.
For example one of the consequences of these studies
is the meaning of the so-called correlation length 
$r_0$ defined by $\xi(r_0)=1$, which is usually considered
to be $r_0\approx 5 \hmp$. In our opinion, due to the fractal properties
identified in this distribution, such a length scale 
is not a characteristic length (nor a ``correlation length'',
neither a length scale which corresponds to the transition
from non-linear to linear structures), but simply a fraction
of the sample's size. 
Future larger samples, like 2dF or SDSS, will permit us to
check these specific properties on larger scales.
However, even beside the question of the possible crossover to homogeneity,
all the structures we see in galaxy distribution have fractal
properties, and hence require a new theoretical framework
for their understanding.

In this lecture we describe our activity 
in the field which includes: (i) data analysis,
(ii) N-body simulations and (iii) 
theoretical modeling.
We refer the interested reader to \cite{slmp98,joyce99a,joyce99b,jampsl00}
for further material on the subject. We also refer to the web
page {\it http://pil.phys.uniroma1.it} where most of
the work we present has been collected.

\section{Complexity}

``More is different''. 
This epochal paper of 1972 by Phil Anderson ~\cite{PWA}  
has set the paradigm   
for what has now evolved  
into the science of Complexity. The idea that ``reality has a   
hierarchical structure in which  
at each stage entirely new laws, concepts, and   
generalizations are necessary, requiring  
inspiration and creativity to just as great a degree   
as in the previous one'' has set a new perspective in  
our view of natural phenomena. The   
reductionist view focuses on the  
elementary bricks of which matter is made, but then   
these bricks are put together in marvelous structures  
with highly structured architectures.   
Complexity is the study of these  
architectures which depend only in part on the nature of   
the bricks, but also have their fundamental laws and  
properties which cannot be deduced   
from the knowledge of the elementary bricks.   
In physical sciences the geometric  
complexity of structures often corresponds to fractal or   
multi-fractal properties ~\cite{MAN}.  
It is not clear whether this is an intrinsic unique property or it   
is due to the fact that we can only  
recognize what we know. May be in the future we shall   
see much more but for the moment one of the elements we  
can identify in complexity is   
its fractal structure.   
Considering then the dynamical  
processes often associated with complex structures we   
have as basic concepts: chaos, fractals,  
avalanches ~\cite{BAK} and 1/f noise. 
Often complex structures arise from  
processes which are strongly out of equilibrium and   
dissipative. There is a broad field, however, which  
is in between equilibrium and   
non-equilibrium phenomena. This  
is the field of glasses and spin glasses which leads to   
highly complex landscapes and to the concept  
of frustration ~\cite{MEZ}.  
Finally an important field  
in which these ideas can also be applied is that of adaptation   
via evolution which is characterized by a degree of  
self-organization and a critical   
balance between periods of smooth  
evolution and dramatic changes ~\cite{KAU}.  
 
We have been working for some years \cite{slmp98} on the  
characterization of galaxy large scale structures and we have found that 
these are fractal in a certain range of scales. A possible crossover 
towards homogeneity is not yet identified  and it is a matter of a wide 
debate \cite{rees99,chown99}. However, no matter the possible 
value of the crossover, the structures observed in three dimensional 
samples are fractal and they require new methods   
both for the techniques of   analysis and for the theoretical  
interpretation and understanding.

\section{Scale Invariant Structures}
 
\begin{figure}[tb]
\epsfxsize 10cm
\centerline{\epsfbox{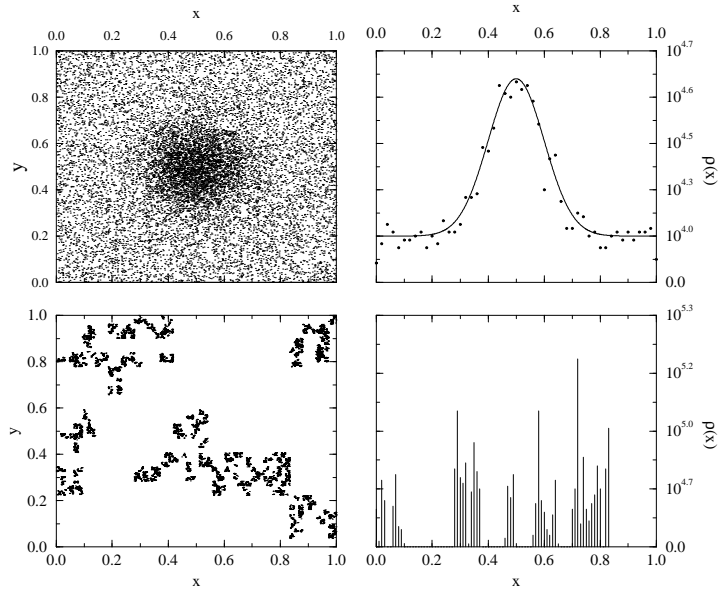}}
\caption{\label{fig1}
Example of analytical and non-analytic structures. Top panels:
(Left)  A cluster in a homogeneous distribution. (Right)
Density profile. In this case the fluctuation
corresponds to an enhancement
of a factor 3 with respect to the average density.
Bottom panels: (Left) Fractal distribution 
in the two dimensional Euclidean space. (Right) Density profile. In this 
case the fluctuations are non-analytical and there is no 
reference value, i.e. the average density. The 
average density
scales as a power law from any occupied point of the structure.
(From Sylos Labini et al., 1998 Phys.Rep. 293, 66)
}
\end{figure}

In order to better define what is an ``irregular structure'' 
let us briefly discuss the properties 
of a regular one (see Fig.\ref{fig1}) 
An  analytical distribution of points is  
characterized by a  
small scale granularity which turns, at larger   
scales, into a well defined background density 
with specific structures corresponding to   
local over-densities (or under-densities).
Let us consider a simple example of a single structure 
(over-density) super-imposed on a uniform background.  
Such a simple  
structure can be characterized by its position, size   
and intensity. One can also define a  
density profile along a line:  
This profile can be well approximated  
by a smooth (analytical) function, which for example  
can be  a constant plus a Gaussian function. If we  
consider the dynamical evolution of our structure   
including the specific interactions  
between its constituent points, we can write a   
differential equation for the smooth  
function of the density profile. In this perspective the   
structure is essentially represented 
 by the three elements: position, size and intensity   
(amplitude). The typical result of 
this study is to understand whether the structure moves, if   
it becomes more or less extended or 
more or less intense. This is the traditional approach   
to the study of structures based on 
the implicit assumption of regularity or analyticity   
which has been the one adopted in  
Statistical Physics before the advent of Critical   
Phenomena in the seventies.  
 
Instead for the case of a strongly irregular  
structure, like for example a simple fractal distribution,  
all the  concepts used to characterize the  
previous picture loose their meaning. There is no   
background density, there are structures  
in many zones and at various scales but it is not   
possible to assign them a specific size  
or intensity. The   
density profile   
is highly irregular at any scale.  
In order to   
give a proper characterization  
of the properties of this structure, one has to look at it   
from a new perspective. A structure  
which consists, for example,  
of a simple stochastic fractal  has   
its regularity in the {\it scale transformation}.  
This naturally leads to power law correlations   
characterized by an exponent, the fractal  
dimension. Also from a theoretical point of   
view, the understanding of the origin of  
the irregular or fractal properties cannot arise   
from the traditional differential equation  
approach but it requires new methods of the type   
of the renormalization group  ~\cite{MAN,EVER,VES}.

\section{Physics of scale-invariant and complex systems}  
 
The physics of scale-invariant and complex  
systems is a novel field which is including   
topics from several disciplines ranging from  
condensed matter physics to geology,   
biology, astrophysics and economics.  
This widespread inter-disciplinarity corresponds to   
the fact that these new ideas allow us to  
look at natural phenomena in a radically new and   
original way, eventually leading to unifying  
concepts independently of the detailed   
structure of the systems.  
The objective is the study of complex, scale-invariant structures,   
that appear both in space and 
time in a vast variety of natural phenomena. New types of   
collective behaviors arise and their understanding represents one 
of the most challenging   
areas in modern statistical physics.  
 
The activity in this field (see e.g the web page of the EC Network  
of  "Fractal structures and self-organization" \cite{network}) results in  
a cooperative effort of numerical simulations, analytical   
and experimental work, and  it  can be characterized by to the following   
three levels: 
  
\begin{itemize} 
 
\item {\it (i) Mathematical or geometrical level.} 
 
This consists in applying the methods o 
f fractal geometry into new areas to get new   
insights into important unresolved  
problems and contribute to a better overall   
understanding. Such an approach   
permits to include into the scientific areas many phenomena   
characterized by intrinsic irregularities  
which have been previously neglected because of   
the lack of an appropriate framework for  
their mathematical description. The main   
examples of this type can be found in the 
 geophysical and astrophysical data.  
 
\item {\it(ii) Development of physical models:  
The Active principles for the generation of Fractal   
Structures.}  
 
Computer simulations represent an essential 
 method in the physics of complex and  
scale-invariant systems. A large number of models 
 have been introduced to focus on specific   
physical mechanisms which can lead spontaneously to fractal structures. 
Here we list some of them, which, in our opinion, represent  
the active principles for processes which generate scale invariant  
properties based on physical processes. In Ref.~\cite{EVER} 
one can find many papers on models like  
Diffusion Limited Aggregation and the Dielectric Breakdown Model.
These models are the prototype of the  
so-called  fractals in which an iteration   
process based on Laplace equation leads 
spontaneously to very complex structures. 
The concept of self-organization is common  
to all the models discussed here but it has   
been especially emphasized in relation to the sandpile model.  
In addition,  to these simplified models  
we know that fractal structures are   
naturally generated in fluid turbulence  
as described by Naiver-Stokes Equations as the fractional   
portion of space in which dissipation actually occurs.  
Also the studies of gravitational  
instabilities suggest that gravity with random initial   
conditions may be enough to generate fractal clustering
(see below).  
Up to now, however, the connections between the two important 
problems   
of turbulence and gravitational clustering and the above 
simplified models are only indirect. Each phenomenon and model  
mentioned seems to belong to a different universality class. 
 
\item {\it (iii) Development of theoretical understanding}  
 
At a phenomenological level scaling theory,  
inspired by usual critical phenomena, has   
been successfully used. This is essential for the  
rationalization of the results of the   
computer simulations and experiments.  
This method allows us to identify the relations   
between different properties and to focus on the  
essential ones. From the point of view of   
the formulation of microscopic fundamental 
 theories the situation is still in evolution.   
With respect to usual equilibrium statistical  
mechanics these systems are far from   
equilibrium and their dynamics is  
intrinsically irreversible. This situation does not seem   
to lead to any sort of ergodic theorem and the temporal  
dynamics has to be explicitly   
considered in the theory \cite{VES}. This,  
together with the concept of self-organization, as   
compared to criticality, represent the main new  
elements for the formulation of   
microscopic theories.   
  
\end{itemize} 

\section{Galaxy Structures and Correlations}

The existence of large scale structures (LSS)    
and voids in the distribution of galaxies    
up to  several hundreds Megaparsecs is well known for twenty years    
\cite{huchra,tully}. 
The relationship among these structures    
on the statistics of galaxy distribution    
is usually inferred by applying the     
standard statistical analysis as introduced    
and developed by Peebles and coworkers \cite{pee80}.    
Such an analysis {\it assumes} implicitly that     
the distribution is homogeneous at very small scale    
($\lambda_0 \approx 5 \div 10 \hmp$). 
Therefore the system is characterized as having small fluctuations     
about a finite average density.   
If the galaxy distribution had a fractal nature 
the situation would be completely different.     
In this case the average density in finite samples is not 
a well defined quantity: it is strongly sample-dependent 
going to zero in the limit of an infinite volume.      
In such a situation it is not meaningful to study    
fluctuations around the average density extracted from sample data.    
The statistical properties of the distribution should   
then be studied in a completely    
different framework than    
the standard one. We have been working    
on this problem since some time   \cite{slmp98}  
by following the original ideas of Pietronero \cite{pie87}.    
The result is that galaxy structures are    
indeed fractal up to tens of Megaparsecs \cite{joyce99a}.    
Whether   a crossover    
to homogeneity at a certain scale  $\lambda_0$,     
occurs or not (corresponding to the absence    
of voids of typical scale larger than $\lambda_0$)      
is still  a matter of debate  \cite{rees99}.    
At present, the problem is basically that the available red-shift surveys    
do not sample scales larger than $50 \div 100 \hmp$    
in a wide portion of the sky and in a complete way. 

Note that Gerard de Vaucouleurs \cite{devac70}
has been the first 
who has considered 
a possible hierarchical structure of  galaxy clustering,
which also implies  a different interpretation of galaxy counts:
{\it ``When inhomogeneities are considered (if at all)
they are treated as unimportant fluctuations 
amenable to first order variational treatment. 
Mathematical complexity is certainly an understandable
justification, and economy or simplicity of hypotheses
is a valid principle of scientific methodology:
but submission of all assumptions to the test
of empirical evidence is an even more compelling law
of science''}. He has related the presence of large 
scale structures to the power-law behavior of the 
(conditional) average density and then to a non-Euclidean exponent
in the number counts as a function of magnitude.

\subsection{Self-Organized Criticality in Self-Gravitating systems} 
 
 The  
 clustering of matter in the universe  
 is hence an important example of the fields in which  
 scale invariance has been  
 observed as a common and basic feature.  
 However, the fact that certain structures  
 exhibit fractal and complex properties does not tell us why this happens.  
 A crucial point to understand is therefore the origin of the general  
 scale-invariance of in the gravitational clustering  phenomenon.  
 This would correspond to the understanding of the origin of  
 self-gravitating fractal structures  
 and of the properties of Self-Organized Criticality (SOC) from  
 the knowledge of the microscopic physical processes at the basis of 
 this phenomenon: Most of 
 the scale free phenomena observed in nature are  
 self-organized,  
 in the sense that they spontaneously 
 develop from the generating dynamical process.   
 Such a project requires a close interaction between three  
 different lines, (i) Data Analysis, (ii) N-body simulations,  
 (iii) Formulations 
 of simple physical models, the first steps towards a real theoretical 
 understanding. 
 Let us see these three points in more detail.   
 
\subsection{Data analysis} 
 
Nowadays there is a general agreement about the fact  
that galactic structures are fractal up to a distance  
scale of $ \sim 30 \div 50 \hmp$ \cite{slmp98,joyce99a} and the   
increasing interest about the fractal versus   
homogeneous distribution of galaxy in the last year  
 \cite{coles98,scara98,rees99,cappi98,martinez99, 
hutton99,chown99,landy99}  
has mainly focused  
on the determination of the homogeneity  
scale $\lambda_0$ 
(See the web page  
{\it http://pil.phys.uniroma1.it/debate.html}  where   
all these materials have been collected). 
The main point in this discussion  
is that galaxy structures are fractal no matter what is 
the crossover scale, and this fact has never been 
properly appreciated. Clearly, qualitatively different 
implications are related to different values 
of $\lambda_0$, which could be possibly found in the new  
galaxy three-dimensional samples which will be 
completed in the next few years. 
From the point of view of data analysis we may identify  
different problems which must be addressed for a  
correct understanding of galaxy structures.

\subsubsection{Characterization of scaling properties}  
 
Given a distribution of points,  
the first main question concerns the possibility 
of defining a physically meaningful average density. 
In fractal-like systems such a quantity depends 
on the size of the sample, and it does not 
represent  a reference value, as in the case 
of an homogeneous distribution.  
Basically a system cannot be homogeneous 
below the scale of the maximum void 
present in a given sample.  
However the complete statistical characterization 
of highly irregular structures is the objective of 
Fractal Geometry \cite{MAN}.

The major problem from the point of view  
of data analysis is to use statistical methods  
which are able to properly characterize scale  
invariant distributions, and hence which are  
also suitable to characterize an eventual  
crossover to homogeneity.   
Our main contribution \cite{slmp98},  
in this respect, has been to clarify that the usual   
statistical methods, like correlation function,  
power spectrum, etc. \cite{pee80},  
are based on the assumption of   
homogeneity and hence are not appropriate  
to test it. Instead, we have introduced and developed  
various statistical tools which are able  
to test whether a distribution is   
homogeneous or fractal, and to correctly characterize  
the scale-invariant properties.  
Such a discussion is clearly relevant also  
for the interpretation of the properties  
of artificial simulations. The agreement about   
the methods to be used for the analysis of future  
surveys such as the Sloan Digital Sky Survey (SDSS)  
and the two degrees Fields (2dF) is clearly a fundamental   
issue \cite{slmp98}. 
 
Then, if and only if the average density is found to be 
not sample-size dependent, one may study the  
statistical properties of the fluctuations  
with respect to the average density itself. 
In this second case one can study  
basically two different length scales. 
The first one is the homogeneity scale ($\lambda_0$), 
which  defines 
the scale beyond which the density fluctuations 
become to have a small amplitude with respect to the 
average density ($\delta \rho < \rho$). 
 The second scale is related to the  
typical length scale of the structures of  
the density fluctuations, and, according to the terminology 
used in statistical mechanics \cite{perezmercader}, it is called  
correlation length $r_c$. Such a scale has nothing 
to do with the so-called "correlation length"  
used in cosmology and corresponding to 
the scale $\xi(r_0)=1$\cite{pee80}, which is instead 
related to $\lambda_0$ if such a scale 
exists. Such a confusion being at the origin of the  
misinterpretation of the concept of clustering in  
modern cosmology \cite{slmp98}.  
 
\subsubsection{Fluctuations}  
 
In the characterization of scaling properties, 
one would like to determine other statistical quantities 
beyond the fractal dimension. Such a global parameter 
is in fact the first one to be determined, but then fractals 
with the same dimension can have completely different 
morphological properties (higher order correlations). 
One point to be studied is the identification 
and characterization of some relevant global quantities 
such as porosity, lacunarity and three point correlation function, 
which are poorly studied in the general case of  
mathematical fractal, and never considered in  
the studies of large scale structures.  
Another possibility  \cite{gsl2001} concerns 
the study   of      
fluctuations      
around the average counts as a function of scale
and we have developed tests   
to study of galaxy distribution both in       
red-shift and  magnitude space. Briefly,    
fluctuations in the counts of galaxies, in a  fractal distribution,     
are of the same order of the average number at all scales     
as a function of red-shift and magnitude.   
For the case of an homogeneous distribution  
fluctuations are instead exponentially or power-law damped.   
We point out that the study of these kind of fluctuations     
can be a powerful test to understand     
the nature of galaxy clustering at very large scales as 
these analysis   can be performed on both 
photometric and redshift galaxy catalogs. It is worth to notice that 
one of the anomalous statistical
properties of critical systems, characterized
by power-law 
long-range correlations systems is that, whatever their size,
they can
never be divided into mesoscopic regions that are statistically
independent. As
a result they do not satisfy the basic criterion of the central limit
theorem
and one should not necessarily expect global, or spatially averaged
quantities
to have Gaussian fluctuations about the mean value \cite{nico}.
The probability density function (PDF) of a global measure in a large
class of highly correlated systems is then 
strongly non-Gaussian. The measurement of such a quantity in
galaxy data is then very interesting to understand the 
PDF of galaxies and its possible relation to other
critical systems.

\subsubsection{Implication of the fractal structure up  
to scale $\lambda_0$} 
  
 The fact that galactic structures  
are fractal, no matter what is the homogeneity scale $\lambda_0$,  
 has deep implication on the interpretation  
of several phenomena such   
as the luminosity bias, the mismatch  
galaxy-cluster, the determination of the average   
density, the separation of linear and   
non-linear scales, etc.,   
and on the theoretical concepts used  
to study such properties \cite{slmp98}.  
 An important point is then to consider the main consequences  
of the power law behavior of the galaxy number 
density, by relating various cosmological parameters
(e.g. $r_0$, $\sigma_8$, $\Omega$, etc.)  
to the length scale $\lambda_0$ \cite{joyce2001}.  
This has been partially done, but a more 
complete picture is still lacking. 
We also note that the properties  
of dark matter are inferred from the ones of visible  
matter, and hence they are closely related.  
If now one observes different   
statistical properties for galaxies and clusters,  
this necessarily implies a change of perspective  
on the properties of dark matter.   
For example  
in most direct estimates of the mass density (visible or dark) of the  
Universe, a central input parameter is the luminosity 
density of the Universe. We have considered \cite{joyce2001} the measurement 
of this luminosity density from red-shift surveys, as a function 
of the yet undetermined characteristic scale $\lambda_0$ at which the  
spatial distribution of visible matter tends to a well defined 
homogeneity. Making the canonical assumption that the cluster mass  
to luminosity ratio ${\cal M}/{\cal L}$ is the universal one, we can  
estimate the total mass density as a function  
$\Omega_m(R_H,{\cal M}/{\cal L})$. Taking the highest estimated cluster  
value ${\cal M}/{\cal L} \approx 300 h M_{\odot}/L_{\odot}$  
and a conservative lower limit $R_H \gtapprox 20 \hmp$, 
we obtain the upper bound $\Omega_m \ltapprox 0.1$ . Note that  
for values of the homogeneity scale $\lambda_0$ in the range   
$\lambda_0 \approx (90 \pm 45) h$Mpc, the value of  
$\Omega_m$ may be compatible with the nucleosynthesis inferred  
density in baryons \cite{joyce2001}. From this perspective one of the  
main arguments used as an indirect evidence of non-baryonic dark matter  
fails, and one has no need to invoke an unknown kind of matter to reconcile 
the observed amount of matter in galaxy clusters with the limits 
of primordial nucleosynthesis  (e.g. \cite{bahacall}).

\subsubsection{Determination  
of the homogeneity scale $\lambda_0$}  
  
This is, clearly, a very important point   
at the basis of the understanding  
of galaxy structures and more generally  
of the cosmological problem. We distinguish  
here two different approaches: direct tests and   
indirect tests. By direct tests, we mean the determination  
of the conditional average density in three dimensional surveys,  
while with indirect tests we refer to other possible analysis,  
such as the interpretation of angular surveys, the   
number counts as a function of magnitude or of distance  or,  
in general, the   
study of non-average quantities, i.e. when the fractal 
dimension is estimated without making an average  
over different observes (or volumes). 
While in the first case one is able to have a   
clear and unambiguous answer from the data, in the second  
one is only able to make some weaker   
claims about the compatibility of the data  
with a fractal or a homogeneous distribution.  
For example   the paper of Wu et al. ~\cite{rees99} 
mainly concerns with compatibility arguments, 
rather than with direct tests. 
However, also in this second case, it is possible to understand   
some important properties of the data, and to  
clarify the role and the limits of some underlying   
assumptions which are often used without a   
critical perspective.  
Clearly the availability of new three dimensional galaxy samples 
in the next few years 
would allow one to study larger volume of space with a better 
statistics, and, possibly, to determine the homogeneity scale.

\subsubsection{Crossover towards homogeneity and Finite size effects} 
 
A related and important point under consideration  
concerns  the correct modeling 
of the possible crossover towards homogeneity.  
If the average density will be ultimately defined  
one would like to properly describe the  
transition from a system with large fluctuations  
(fractal) to a distribution with small fluctuations 
(homogeneous with small amplitude and correlated fluctuations). 
A number of statistical tools (correlation function, 
power spectrum, etc.) can be useful in this respect,  
but one has to correctly understand some subtle 
properties due to finite size effects.  
For example, in a finite sample, the power spectrum 
will always show a maximum followed by a decay  
(for $k \rightarrow 0$): such a break is due to a finite size 
effect related to the determination of the average  
density inside the sample itself. 
This has been often and incorrectly 
associated with a real change of the correlation properties 
of the distribution. 
Even in the case of a smooth distribution 
the standard methods used for the characterization 
of correlation must be carefully revised. 
This also particularly interesting for  
the analysis of cosmological N-body simulations which  
indeed show a smooth transition from small scale 
fractality to large scale homogeneity. 

\section{N-body simulations} 
 
We have started to study the problem of the  
self-gravitating gas in a periodic volume. 
We have used high resolution N-body simulations to study the  
dynamical evolution of a gas of particles, 
initially distributed according to Poisson  
statistics, with periodic boundary conditions, 
and (for the moment) without the effect of 
space expansion. The aim of this project is to 
understand first a simple case of  
clustering process to then study more  
sophisticated simulations which involve  
space expansion and a particular choice of initial conditions.  
The results, which must be tested with large  
simulations as   the number of particles 
used is in the range $10^3 \div 10^4$, is that 
the system spontaneously develops 
self-similar fluctuations, characterized 
by a fractal dimension $D\approx 2$. 
There is a list of new type of question which we would like to address:
Can gravity develop a critical equilibrium ? Is the fractal
dimension $D \approx 2$ a characteristic exponent of gravity ? 
How long in time and how deep in space does the critical 
behavior extend ? 
Basically, in the standard picture described by the continuous
equations one has a linear or non-linear amplification of 
smooth fluctuations. In the new perspective there is a transfer
of non-analytic clustering (granulosity) from small to large scales.
The continuous fluid description simple neglects the effect due
to small scale granulosity.

There are large N-body cosmological simulations 
which are publicly available.  
The clustering in these simulations is  
due to a combination of effects (space expansion,  
initial conditions, properties of dark matter and  
gravitation) and hence it is more difficult 
to understand the influence of each of these  
effects at a time. However  
it is interesting to consider, as a first step, 
the statistical properties of both initial and  
final conditions in these simulations.  
We have studied the statistical properties of 
cosmological N-body simulations based on CDM-like models, showing that  
they develop fractal structures almost independently 
on a wide choice of   
initial conditions and cosmological parameters. 
In such a case, however, the fractal extends in a 
relatively small range of scales (i.e. $0.1 \div 20 \hmp$) 
and a crucial point in this respect is the fact, 
that self-similar fluctuations require a long 
time to develop over a large range of scales  
(up to $\sim 100 \hmp$ or more) from Gaussian initial conditions. 
 
A related question concern the implementation of  
the initial conditions of N-body simulations.  
In standard models (like CDM's) the initial conditions  
are due to a combination of properties  
of the quantum fluctuations of the early universe and  
the specific properties of the considered dark matter. 
However one has predictions on the initial 
continuous density field and its correlation 
properties. 
How to discretize the initial continuous density field ? 
The answer to this basic question is clearly
fundamental in order to relate the properties
of the initial continuous density field to the 
the properties of a discrete set of points, with
which the initial conditions of N-body simulations are
usually set up.
 
\section{Interpretation and Modeling}  
 
\subsection{Exponents versus amplitudes}

From the theoretical point of view, 
the only relevant and meaningful quantity is the  exponent 
of the power law correlation function (or of the space density), 
while the amplitude of the correlation  function, or of the space 
density, is just  related to the sample size and to 
the lower cut-offs of the distribution.  The geometric 
self-similarity has deep implications for the  non-analyticity 
of these structures. Indeed, analyticity or regularity would 
imply that at some small scale  the profile becomes smooth 
and one can define  a unique tangent. Clearly this is impossible 
in a self-similar structure because at any small scale a new 
structure appears and the  distribution is never smooth.
 Self-similar structures are therefore intrinsically irregular 
at all scales and correspondingly one  has to change the 
theoretical framework into one which is  capable of 
dealing with non-analytical fluctuations. This means
 going from differential equations to something like 
the  Renormalization Group to study the exponents.
For example the so-called ``Biased theory of galaxy
formation'' \cite{kai84}
is implemented considering 
the evolution of  density fluctuations within an analytic 
Gaussian framework,  while the non-analyticity of fractal 
fluctuations  implies a breakdown of the central limit 
theorem which is the  cornerstone of Gaussian processes 
\cite{pie87,VES,slmp98}.

\subsection{Fractal cosmology in an open universe}

The clustering of galaxies is well characterized by fractal properties, 
with the presence of an eventual cross-over to homogeneity still a matter 
of considerable debate.  We have discussed and considered  
the cosmological implications of a fractal distribution of 
matter, extending to an arbitrarily large scale \cite{jampsl00}.  
Such an  open model of universe 
can be treated consistently within the framework of  
the expanding universe solutions of Friedmann, with the fractal 
being a perturbation to an open cosmology in which the leading  
homogeneous component is the cosmic microwave 
background radiation (CMBR).  
This new type of cosmology, inspired by the observed galaxy 
distributions, provides a simple explanation for the recent  
data which indicate the absence of deceleration 
in the expansion ($q_o \approx 0$).  
Moreover the `age problem' is essentially eliminated. 
The model leads to a new scenario 
for the explanation of the observed isotropy of the CMBR. The  
radiation originates mostly in the annihilation processes which leave 
behind the large voids, with the residual fractal matter leading 
to small perturbations. Nucleosynthesis and the formation of 
structure can also be addressed in this new framework.

\subsection{Force Distribution}

One of the main properties it is possible to calculate  
in the analysis of the gravitational characteristics of  
a poissonian distribution of points, is the  
probability distribution of the (Newtonian) gravitational 
force. Such a distribution is known as the  
Holtzmark distribution \cite{chandra}. 
We have  considered  the generalization of the Holtzmark  
 to the case of  
a fractal set of sources \cite{gslp98}. 
We have shown that, in the case of  
real structures in finite samples,  an important role 
is played by morphological properties 
and finite size effects. 
For dimensions smaller than $d-1$  
(being $d$ the space dimension) 
the convergence of the net gravitational 
force is assured by the fast decaying of the density, 
while for fractal dimension $D>d-1$ the morphological 
properties of the structure  
determine the possible convergence of the force 
as a function of distance.  
The relashionship between peculiar velocity and gravitational 
fields has been considered in  \cite{jampsl00}.

\section{Conclusions}

 The  
 clustering of matter in the universe  
 is hence an important example of the fields in which  
 scale invariance has been  
 observed as a common and basic feature.  
 However, the fact that certain structures  
 exhibit fractal and complex properties does not tell us why this happens.  
 A crucial point to understand is therefore the origin of the general  
 scale-invariance of in the gravitational clustering  phenomenon.  
 This would correspond to the understanding of the origin of  
 self-gravitating fractal structures  
 and of the properties of Self-Organized Criticality (SOC) from  
 the knowledge of the microscopic physical processes at the basis of 
 this phenomenon: Most of 
 the scale free phenomena observed in nature are  
 self-organized,  
 in the sense that they spontaneously 
 develop from the generating dynamical process. 
 For example some interesting attempts to understand why 
 gravitational clustering generates scale-invariant structures have been
 recently proposed by de Vega et al. \cite{devega1,devega2,devega3}.
 Basically, the Physics should shift from the study of {\it "amplitudes"}
 towards {\it "exponents" }
and the methods of modern Statistical Physics should be adopted.
This requires the development of constructive interactions 
between the two fields.

\begin{acknowledgments}
We thank Y.V. Baryshev, R. Durrer, P.G. Ferreira, A. Gabrielli, M. Joyce,    
and M. Montuori for useful discussions.    
This work is partially supported by the      
 EC TMR Network  "Fractal structures and  self-organization"       
\mbox{ERBFMRXCT980183} and by the Swiss NSF.      
\end{acknowledgments}

\end{document}